\documentclass{iopconfser}
\usepackage{graphicx}
\usepackage{amsmath}
\usepackage{xcolor}  
\usepackage{soul} 
\usepackage[table]{xcolor}
\usepackage[left=2.5cm,right=2.5cm,top=4cm,bottom=2.7cm]{geometry}
\begin{document}

\title{MILP-driven Network Planning Framework for Energy Efficiency and Coverage Maximization in IoT Mesh Networks }

\author{Ishmal Sohail$^{1}$, Attiq Zeeshan$^{1}$, M. Umar Khan$^{1}$, Syed Zubair$^{2}$, Rana Fayyaz Ahmad$^{3}$, Faizan Hamayat$^{3,4}$}

\affil{}
\affil{$^1$Department of Computer Engineering, COMSATS University Islamabad, Pakistan}
\affil{$^2$Military Technological College, Muscat, Oman}
\affil{$^3$Artificial Intelligence Technology Centre, Islamabad, Pakistan}
\affil{$^4$Shenzhen AiMall Technology Co. Ltd., Shenzhen, China}

\email{FA22-BCE-040@isbstudent.comsats.edu.pk, FA22-BCE-028@isbstudent.comsats.edu.pk, umar\_khan@comsats.edu.pk, szgilani@gmail.com, fayyaz.ahmad@ncp.edu.pk, and faizan.hamayat@ncp.edu.pk}

\begin{abstract}
In the era of digital transformation, the global deployment of internet of things (IoT) networks and wireless sensor networks (WSNs) is critical for applications ranging from environmental monitoring to smart cities. Large-scale monitoring using WSNs incurs high costs due to the deployment of sensor nodes in the target deployment area. In this paper, we address the challenge of prohibitive deployment costs by proposing an integrated mixed-Integer linear programming (MILP) framework that strategically combines static and mobile Zigbee nodes. Our network planning approach introduces three novel formulations including boundary-optimized static node placement (MILP-Static), mobile path planning for coverage maximization (MILP-Cov), and movement minimization (MILP-Mov) of the mobile nodes. We validated our framework with extensive simulations and experimental measurements of Zigbee power constraints. Our results show that boundary-optimized static placement (MILP-Static) achieves 53.06\% coverage compared with 33.42\% of random approach. In addition, MILP-Cov for path planning reaches 97.95\% coverage while movement minimization (MILP-Mov) reduces traversal cost by 40\%. Our proposed framework outperforms the benchmark approaches to provide a foundational solution for cost-effective global IoT deployment in resource constrained environments.           
\end{abstract}

\section{Introduction}
The strategic planning of communication networks is paramount for cost-effective deployments in the era of digital transformation. While fifth-generation (5G) cellular network leverage data-driven dimensioning, clustering, and optimization to balance coverage, capacity, cost for diverse use cases, analogous challenges plague large-scale WSNs and Iot Networks. Specifically, the prohibitive expense of deploying static sensor nodes across vast geographical area remains a critical barrier to scale environmental monitoring and smart city applications. Existing methodologies, including radio network dimensioning (RND) with heuristic optimization \cite{net3}, traffic-based base station clustering \cite{net2}, and mixed integer linear programming (MILP) driven cellular topology design \cite{net1}, focus primarily on cellular infrastructure, overlooking the unique resource-constrained WSN/IoT deployments. In this aspect, this work contains an integrated MILP formulation that rethinks cost efficiency through the novel co-optimization of static and mobile Zigbee nodes, addressing coverage, power, and movement constraints in resource-constrained IoT environments.  

The IoT nodes for wireless communication usually contain a Zigbee module for the connectivity and data sharing between each node. Zigbee module transfers the data through a wireless medium that is generally an internet connection or a WIFI communication. The Zigbee-based IoT nodes can be categorized into static IoT nodes and mobile IoT nodes. The terms static and mobile refer to IoT nodes with a fixed position for data collection from a fixed area or is movable, enabling it to collect data from different regions of the network. Since the placement of the IoT nodes plays a vital role for connectivity and data transfer, a MILP technique \cite{10225271} is proposed for determining the connections throughout the network. Both Static and Mobile IoT nodes can be placed in the network for data collection and transmission. Another parameter for network planning involves the consumption of resources. In this regard, the system design must be energy efficient as the IoT nodes operate consistently \cite{10510318}. The IoT nodes work on real-time data collection, the energy consumption has become an important parameter for the functioning of IoT nodes. Static IoT nodes require less energy as compared to the mobile IoT nodes due to the continuous movement and data collection from different areas within a network.

This study presents an MILP-based network planning framework that improves energy consumption and coverage within Zigbee-based mesh networks. We propose three MILP models: MILP-Static, MILP-Cov, and MILP-Mov, to optimize the placing of static nodes as well as mobile nodes. The goal is to achieve maximum coverage, stabilize connectivity, and also reduce unnecessary node mobility. Our framework integrates key aspects of network design, including communication protocols, routing strategies, and topology planning. Zigbee is selected as the communication protocol due to its low-power consumption for a 10 to 100-meter range. The proposed models are designed using real-world scenarios and offer a complete solution to planning energy-efficient and resource-efficient IoT networks.

\section{Literature Review}
Network planning in IoT has received attention in earlier studies to enhance energy efficiency, path development, and reliability of communication between nodes. Many researchers are incorporating spatial graph attention and temporal self-attention to enhance the performance of networks \cite{10263625}. For instance, in \cite{10840243}, authors proposed enhanced traffic forecasting TIIDGCN in IoT networking by capturing the multi-scale spatial-temporal dynamic graph learning, which outperforms baseline models on real-world traffic datasets. Energy consumption plays a vital role for planning of an IoT network, since it must be energy-efficient to make the data transfer efficient and cost-effective. To this end, the authors of \cite{9548945} proposed a energy efficient client-side access point approach for WiFi/LiFi networks for efficient energy usage in next-gen IoT networks. Another author in \cite{10411996} proposes a LAQEER protocol for connection reliability and efficient energy consumption within WSNs, which reduces packet loss and latency. This approach performs better than existing protocols in terms of ratio of packet delivery and energy usage. Whereas, an EEMSR protocol is proposed by the authors in \cite{9581300} for IoT networks to optimize multi-hop routing with clustering to reduce communication overhead, which integrates genetic algorithms and analytic hierarchy process for weight assignment and inter-clustering routing in IoT networking. Looking forward to communication protocols for optimal network planning, widely used protocols are LoRaWAN and ZigBee because of their low power transmission across the network. For instance, in \cite{9803130} author optimizes the deployment of the LoRaWAN communication protocol to propose a framework for the building and selection of network models that enhance lifespan, delay, and capacity management. On the other hand, the author in \cite{10146383} developed an antenna-embedded Zigbee WSN for IoT applications in network planning, enabling real-time monitoring and cloud storage integration of network plans. For network topology related to communication protocol, LoRaWAN works with a bus topology, whereas Zigbee is suitable for a mesh topology. A mesh network creates a supreme communication channel between IoT nodes. A wireless mesh network enhances bandwidth, reliability, and power efficiency by providing a detailed view of interconnection methodologies, challenges, and other future opportunities, as in \cite{9656902}.  In past research, many authors proposed their frameworks for the deployment of nodes for full area coverage with minimal resource use for IoT node network planning. To illustrate, in \cite{10553388}, authors proposed the MILP framework for planning the network based on IoT nodes, which results in superior performance than their benchmark frameworks. In this study, we develop a comprehensive MILP-based network planning framework for Zigbee-enabled IoT mesh networks. We introduce three models, including the MILP-Static (for static nodes placement), MILP-Cov (for mobile nodes placement), and MILP-Mov (for mobile nodes movements), for the optimal placement of nodes including static as well as mobile nodes. Our proposed framework improves energy efficiency, maximizes coverage, and reduces unnecessary mobility.

\section{Methodology}
The proposed methodology is structured around a series
of interdependent steps, beginning with identifying key
variables, followed by developing mathematical models, simulation scenarios, and validation against predefined benchmarks. To achieve maximum area coverage with minimal resource use, a novel framework was developed that integrates advanced mathematical modeling, optimization techniques, and simulation-based analysis to ensure the implementation of the proposed solutions.
\vspace{-.5cm}
\subsection{Proposed Framework}
The proposed framework aims to optimize the Zigbee-based static nodes deployment and plan the path of Zigbee-based mobile nodes inside a wireless network to plan an optimal IoT network. The framework integrates MILP formulations involving 
additional constraints, a mathematical model, and a simulation
tools to address the challenges of efficient node placement and
Path planning. 
\vspace{-.5cm}
\subsection{IoT Network Profile}
Assume we have a network with an area divided in the form of grids which contains $M \times $N cells with unit square. Every location of the cell has coordinates which are (i,j) laying between $1 \leq i \leq M$ and  $1 \leq j \leq N$. Begin by positioning the static nodes along the grid walls. Then, path planning of mobile nodes is done for maximal coverage of the area and to reduce overlapping. The goal is to place the static nodes at walls and then plan the path of mobile nodes such that sensor nodes cover all the grid cells within the network area. The symbolic representation of system parameters and variables are given in Table 1.
\renewcommand{\arraystretch}{1}
\begin{table}[h!]
\centering
\caption{Variables and parameters.}
\begin{tabular}{|c|>{\centering\arraybackslash}p{8cm}|} 
\hline
\textbf{Variables} & \textbf{Description} \\
\hline
$x_{i,j}$ & Static node placement indicator. \\
\hline
$x_{l,k,i,j}$ & Mobile node placement indicator. \\
\hline
$c_{i,j}$ & Grid cell coverage status. \\
\hline
$c_{l,k,i,j}$ & Individual node’s cell coverage. \\
\hline
$N_s$ & Total static node count. \\
\hline
$N_m$ & Total mobile node count. \\
\hline
$K_{\text{max}}$ & Maximum number of movement steps. \\
\hline
$r_{s}$& Sensing radius of Zigbee module.\\
\hline
$M$ & Grid height (rows). \\
\hline
$N$ & Grid width (columns). \\
\hline
$d_i, d_j$ & Displacement within movement range. \\
\hline
$\rho_x, \rho_y$ & Mobile node travel range. \\
\hline
overlap\_limit & Maximum allowed overlap per cell. \\
\hline
$\alpha$ & Weight for boundary cell coverage. \\
\hline
\end{tabular}
\end{table}

Static nodes and mobile nodes are indicated by using $N_s$ and $N_m$ variables. A binary decision variable x[i,j] indicates whether a static node is placed in grid cell (i,j), while c[i,j] indicates whether a static node covers the cell (i,j). For mobile nodes, x[l,k, i,j] represents a binary decision variable that indicates whether the mobile node $l^{th}$ is in the cell (i,j) at timestep k, and c[l,k, i,j] shows that cells (i,j)  are covered during timestep k by the mobile nodes. Sensing range of the Zigbee module is $r_{s}$= 1. As shown in Figure 1, we assume that sensing range $r_{s}$ =  1 means 100 meters in a $3 \times 3$ grid is in radius. The movement constraints of the mobile node are determined by $\rho_x$ and $\rho_y$, which limit their horizontal and vertical movement per timestep. $K_{max}$ is maximum steps allowed to every mobile node fo movement.
\begin{figure}
    \centering
    \includegraphics[width=0.50\linewidth]{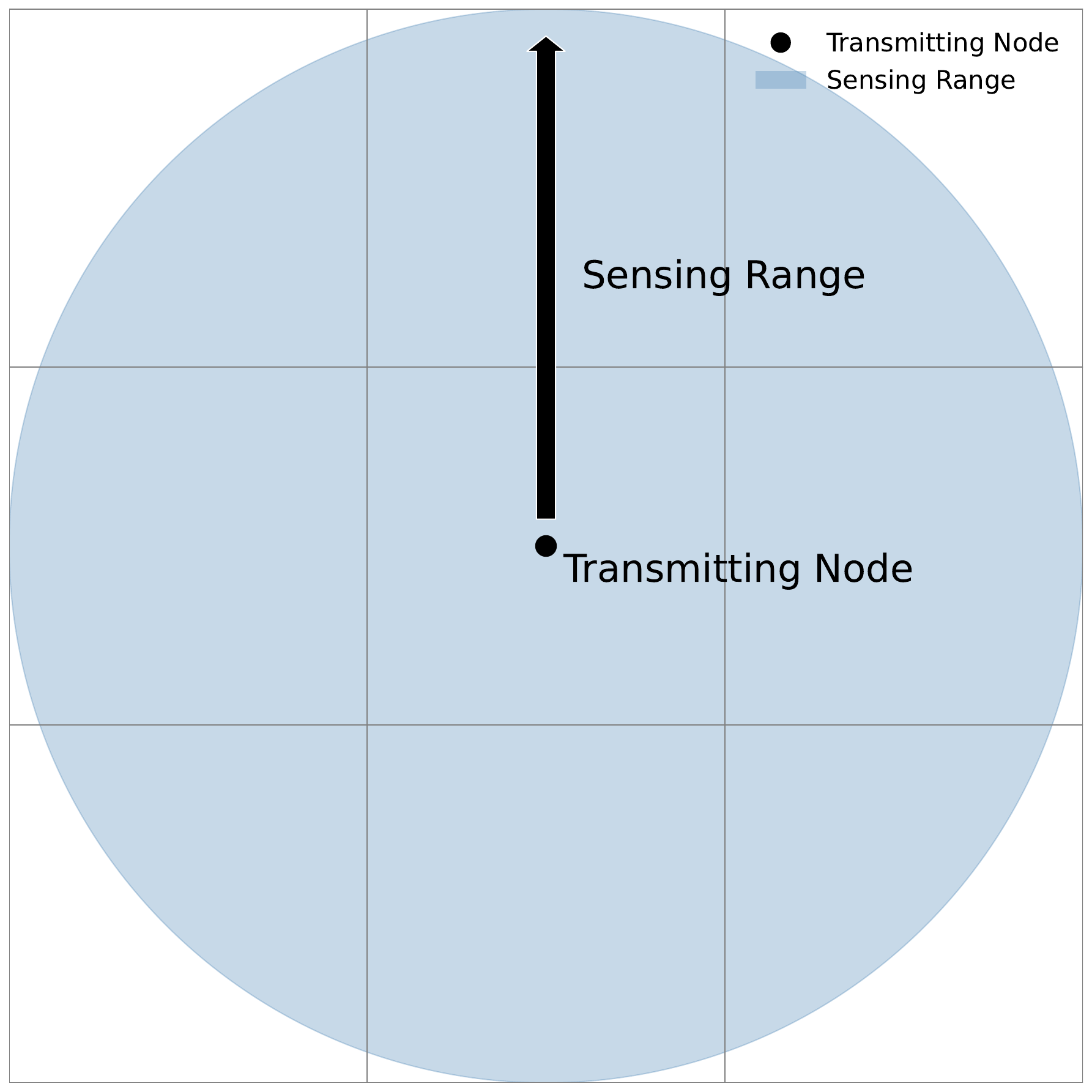}
    \caption{Sensing range of the considered Zigbee module for WSN deployment.}
    \label{fig: 1}
\end{figure}
The overlap between nodes is controlled by a parameter, the overlap limit, ensuring non-redundant coverage in grid cells. Additionally, the static coverage map (i,j) pre-computes the coverage provided by static nodes for each grid cell, and $\alpha$ is a parameter that balances the importance of coverage for interior versus boundary cells within a grid. These variables together allow for an optimization problem aimed at maximizing the area coverage while fulfilling all movement, coverage, and overlap constraints. The example of a network model is given in Figure 2. The various
parameter values for this network are M=7, N=7, $N_s$ = 3, $N_m$ = 2, $K_{max}$ = 3, $r_s$=1, and $ \rho_x = \rho_y = 2$. This is a $7 \times 7$ grid network with 3 static nodes and 2 mobile nodes with Zigbee module installed on them, providing coverage for 49 out of 49 cells. The grid cells not covered by static nodes are covered using two mobile nodes with a one-step traveling range of two cells in each direction.

\section{MILP-Driven System Modeling}
The proposed model employs three MILP formulations to optimize the area coverage and network planning. The system model takes a limited number of static nodes and mobile nodes for network planning, for uncovered cells, for network area coverage, and total trip time. The model contains a parameter for finding the redundant coverage level or the coverage that overlaps while planning the path. Key contributions of the proposed formulations are given below: 
\subsection{MILP-Static: Static Node Placement}
MILP-Static places the static nodes to maximize the overall coverage of the network area by giving importance to covering the boundary cells of the network. The placement of a static node can be done by using the following mathematical formulations.
\begin{figure}
    \centering
    \includegraphics[width=0.50\linewidth]{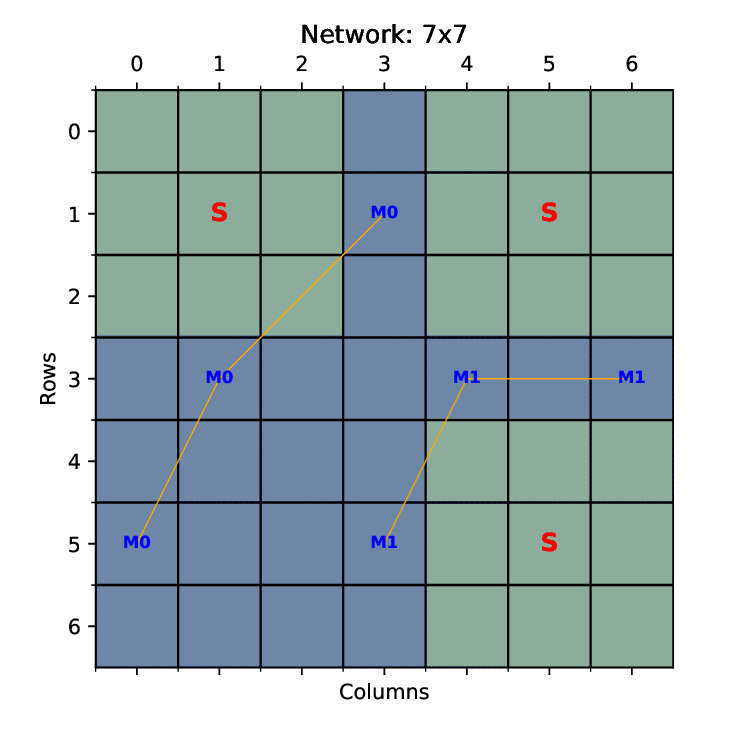}
    \caption{Coverage path for 7x7 network area with 3 static nodes, 2 mobile nodes deployed using MILP-Static and MILP-Cov.}
    \label{fig:enter-label}
\end{figure}
 \begin{equation}
\max\left( \sum_{s=1 \atop (i,j) \in A} c_{i,j}^s 
+ \alpha \sum_{s=1 \atop (i,j) \in A}^{N_s} c_{i,j}^s \right)
\end{equation}
 \begin{equation}
            \text{s.t.}\sum_{(i,j)\in C} x_{i,j}^s =1; \quad s=1, \dots ,N_s
\end{equation}
\begin{equation}
c_{i,j}^s = \sum_{p=-r_s}^{r_s} \sum_{q=-r_s}^{r_s} x_{i+p, j+q}^s\quad ;\forall (i,j)\in C; \;s=1, \dots, N_s
\end{equation}

\begin{equation}
\sum_{s=1}^{N_s} c_{i,j}^s \leq\hat{s}_o; \quad\forall (i,j) \in C
\end{equation}
    \textbf{Objective Function:} The variable $ c_{i,j}^s$ in (1) indicates whether cells (i,j) are covered by static nodes or not. The objective maximizes the total covered cells, with additional weight $ \alpha$ for boundary cells (B), defined as the edges of the grids. Using $\alpha < $ 1 prioritizes boundary cell coverage. 

    \textbf{Position Constraint: } Each sensor node must occupy exactly one grid cell, ensuring no node spans multiple cells, expressed in (2).
   
    \textbf{Coverage Constraint: } A cell (i,j) is covered if it is within sensing range ($r_s$) of a node s. This ensures $c_{i,j}^s = 1$ when (i,j) falls within $r_s$ of s, expressed in (3).
   
    \textbf{Overlapping Constraint:} To prevent resource inefficiency, overlapping coverage of cells is restricted. A cell (i,j) can be covered by at most $\hat{s}_o$ nodes to limit excessive overlap.

    \subsection{MILP-Cov: Coverage Maximization}
        After the placement of the static node by MILP-Static, the mobile will then be traversed by MILP-Cov to cover the uncovered cells $(i,j) \in \overline{C}_s; $ for maximal coverage of the network area. Planning the path of mobile nodes for maximal coverage area takes the form of the MILP-Cov objective function as:
        \begin{equation}
            \max \sum_{(i,j)\in \overline{C}_s} c_{i,j}
        \end{equation}
\begin{equation}
    \text{s.t.} \sum_{(i,j)\in\overline C} x_{i,j}^{l,k}=1; \quad l=1, \dots, N_m;k=1,\dots, K_{\max}
\end{equation}

$  \quad x_{i,j}^{l,k+1} = \sum_{p=-\rho_x}^{\rho_x} \sum_{q=-\rho_y}^{\rho_y} x_{i+p, j+q;}^{l,k} \quad\forall (i,j)\in\overline{C}_s;$

\begin{equation}
     \quad \quad \quad l=1, \dots, N_m; \; k=1, \dots, (K_{\max}-1)
\end{equation}
\\
\textbf{Objective Function:} Maximizing the sum of coverage variables $c_{i,j}$ over uncovered grid cells ${C}_s$, ensuring maximum coverage area by mobile nodes is defined by (5).

\textbf{Position Constraint: } Each mobile node can occupy only one grid cell per time-step, enforced by ensuring the sum of the variable defining position $x_{i,j}^{l,k}$ across all uncovered cells equals 1 at every time-step.

\textbf{Mobility Constraint:} Movement of mobile nodes must e within one cell per time-step (7), with their current position constrained to cells within their one-step traveling range ($\rho_x = \rho_y$) of the previous position. 
\\

      $  c_{i,j}^{l,k}=\sum_{p=-r_s}^{r_s}\sum_{q=-r_s}^{r_s} x_{i+p,j+q;}^{l,k}, \quad\forall (i,j) \in\overline{C}_s;$

\begin{equation}
     \quad \quad \quad l=1, \dots, N_m; \; k =1\dots, K_{\max}
\end{equation}

   $\quad \quad  c_{i,j}\geq c_{i,j}^{l,k}; \quad\forall (i,j) \in\overline{C}_s; \; l=1 \dots,N_m;$ 

\begin{equation}
    k=1,\dots, K_{\max}
\end{equation}
\begin{equation}
    c_{i,j}\leq\sum_{l=1}^{N_m} \sum_{k=1}^{K_{\max}} c_{i,j}^{l,k}; \quad\forall (i,j) \in\overline{C}_s
\end{equation}
\\
        \textbf{Coverage Constraint:} (8) ensures $c_{i,j}^{l,k}$ is equal to 1 when grid cell (i,j) lies inside the mobile node sensing range l-th at time-step k.  (9) ensures that the grid cell covered at time-step k by mobile node $N_m$ should be covered in the remaining cells. (10) sets $c_{i,j}$ to 0 if no mobile node covers the grid cell (i,j) during any time-step k.
\\
\begin{equation}
    \sum_{l=1}^{N_m} \sum_{k=1}^{K_{\max}} c_{i,j}^{l,k} \leq overlap\_limit; \quad \forall (i,j) \in \bar{C}_s
\end{equation}

   $\quad \quad  0\leq c_{i,j}, \; c_{i,j}^{l,k}\leq 1; \quad\forall (i,j) \in \overline{C}_s; \; l=1,\dots, N_m; \; $

\begin{equation}
    \quad \quad \quad k=1, \dots, K_{\max}
\end{equation}

\textbf{Overlapping Constraint:} (11) limits overlapping coverage by allowing each grid cell to be covered at most $\hat{m}_o$  times, which ensures maximal area coverage while permitting overlapping. Setting $\hat{m}_o =1$ may restrict mobile nodes' mobility and reduce coverage. (12) bounds the continuous variables  $c_{i,j}$ and$ c_{i,j}^{l,k}$ between 0 and 1, ensuring their values must not be grater than 1, even many times the grid cells are covered.  

    \subsection{MILP-Mov: Movement Minimization}
        The method of planning the path of a mobile node is to reduce its movements to attain the expected ratio of coverage(cr). This can be done using the given optimization strategy: MILP-Mov.

      \begin{equation}
    \min \sum_{k=1}^{K_{\max}} \sum_{{l=1 \atop (i,j) \in \overline{C}_s}}^{N_m} x_{i,j}^{l,k}
\end{equation}
\begin{equation}
    \text{s.t.}\sum_{(i,j)\in\overline{C}_s} x_{i,j}^{l,k} \leq 1; \quad l=1, \dots, N_m; \; k=1, \dots, K_{\max}
\end{equation}

    $\quad \quad \quad \quad \quad \sum_{(i,j)\in C} c_{i,j}\geq \text{cr}\cdot |C|$

\begin{equation}
    Constraint\quad (7),(8),(9),(10),(11),(12)
\end{equation}
\textbf{Objective Function:} As movement of mobile nodes is the major energy expense, reducing energy consumption by minimizing the total cells visited by mobile nodes. The (13) calculates the total cells covered by all mobile nodes across time-steps k.

\textbf{Position Constraint:} Extends the position constraint in MILP-Cov in (6) to allow mobile nodes to stop moving once the desired coverage area ratio is achieved. This enables energy savings by halting movements when sufficient coverage is achieved.

\textbf{Coverage Ratio Constraint:} Ensures the overall coverage by both mobile and static nodes meets or exceeds the expected ratio $c_r$ . For $c_r=1$, the algorithm aims for 100\% area coverage. Unlike MILP-Cov, the stopping criterion depends on achieving this coverage ratio rather than a fixed $K_{\max}$.

\section{Experiments and Results}
In this section, we detail the findings obtained through the implementation of improved MILP-based IoT network planning, emphasizing trends, patterns, and their significance in the context of achieving maximum coverage area with minimal resource use.

\subsection{Experimental Setup}
For the network area, we consider the $7\times7$ grid having $1\times1$ size for every grid cell. The variation of static nodes lies from 0 to 3, and the mobile nodes vary from 1 to 3. We suppose the sensing range of $r_s$ = 1, mobility range $\rho_x$ as 1 to 5, and overlapping coverage factor for both static and mobile nodes as overlap\_limit=2. The algorithms were implemented in Jupiter Notebook 7.0.8 by Anaconda Navigator in Python (ipykernal). The formulations are performed in DELL with processor Intel(R) Core i7 with 2.80 GHz frequency. The system type was a 64-bit Windows 10 Pro operating system with RAM of 16 GB (15.9 GB usable).

\subsection{Placement of Static Nodes}
To place the nodes in the network area, the first task is the placement of static nodes to target the maximal coverage area. For this, the comparison of the results based on the coverage percentage of two static node placement strategies is done. The two strategies are: Random placement and MILP-Static formulations. Table 2 compares the coverage area achieved using both static node placement strategies, by varying the static as well as mobile nodes for 2 different network areas.
\begin{table}[h!]
    \centering
      \caption{Network coverage by MILP-Cov in comparison with Random and Static approaches.}
      \begin{tabular}{|c|c|c|c|c|}
        \hline
        \textbf{Network Area} & $N_m$ & $N_s$ & \multicolumn{2}{c|}{\textbf{Area Coverage (\%)}} \\ 
        \cline{4-5}
        & & & \textbf{Random} & \textbf{MILP-Static} \\ 
        \hline
        & 1 & 1& 33.42& 53.06\\ & 1 & 2& 49.75& 69.38\\ $7\times7$ & 2 & 1& 65.96& 81.63\\ & 2 & 2& 82.14& 91.83\\ & 3 & 1& 77.17& 95.91\\ & 3 & 2& 85.93& 97.95\\ 
        \hline
        \multicolumn{5}{c}{}\\
    \end{tabular}
    \label{tab:coverage}
\end{table}
We suppose that after static nodes placement, planning the path is done using MILP-Cov for mobile nodes, in terms of area coverage. Static nodes are placed randomly without following any technique or pattern in the Random strategy. The values of the coverage area are taken by averaging over 5 simulation runs with $K_{max}=2$. We observe that the results generated by the MILP-Static are more efficient and better than other strategies. So, the MILP-Static strategy is used in the model to simulate the results along with MILP-Cov and MILP-Mov. Figure 3 
\begin{figure}
    \centering
    \includegraphics[width=0.72\linewidth]{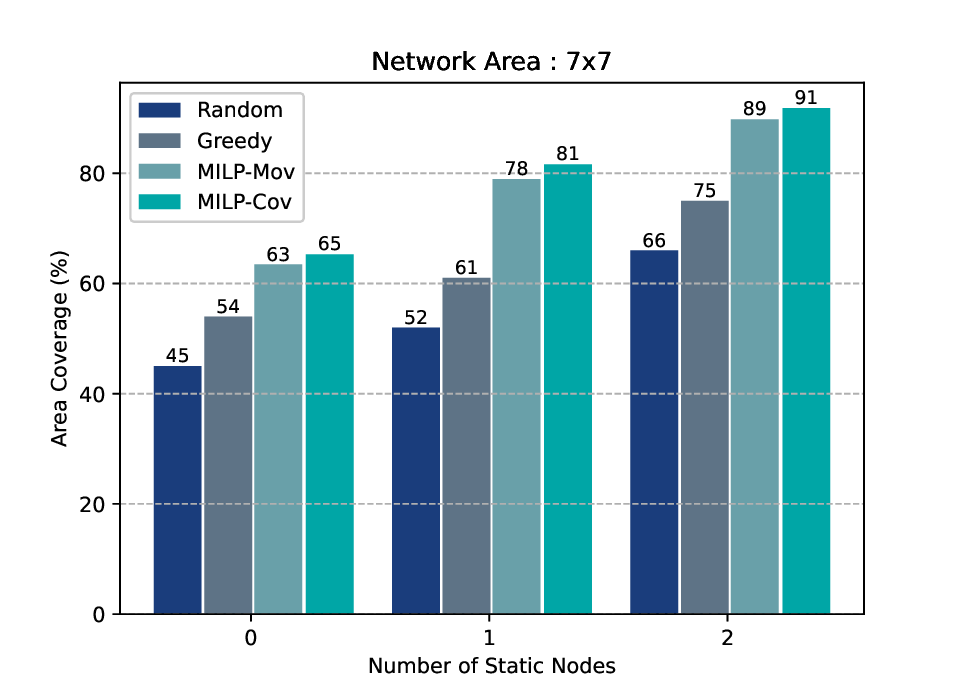}
    \caption{Network coverage area with $N_m=1$ and $K_{\max}=4$ having variable static nodes by MILP-Static.}
    \label{fig:enter-label}
\end{figure}
show results for static node placement by MILP-Static and path planning of Mobile nodes using Random, Greedy, Mov, and Cov strategies. 
\begin{figure}
        \centering
        \includegraphics[width=0.7\linewidth]{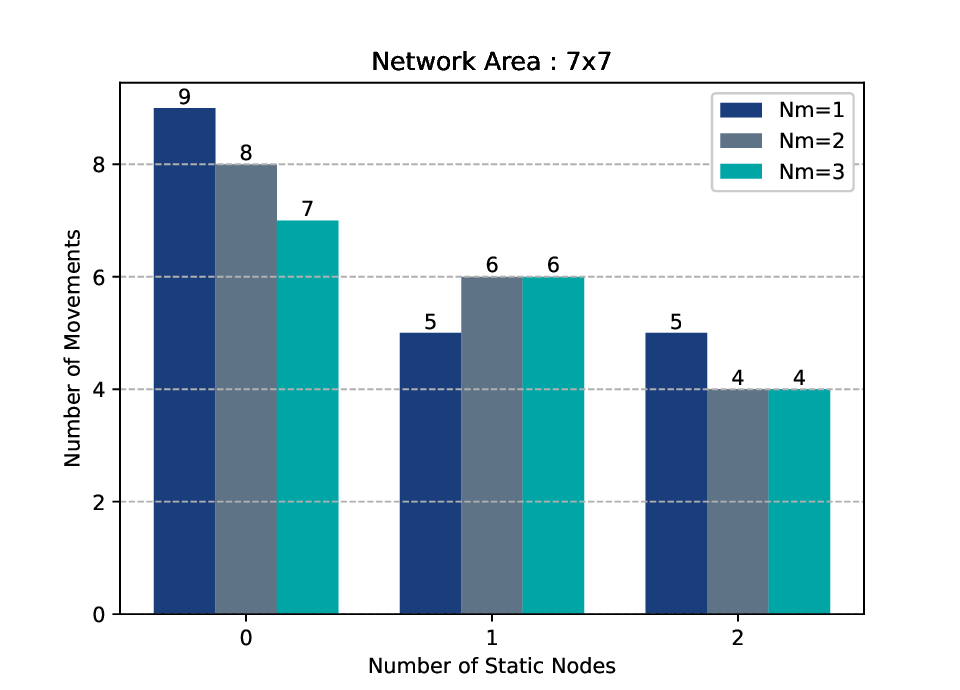}
        \caption{Total movements achieved for full area coverage using MILP-Mov and MILP-Static for mobile and static nodes, respectively.}
        \label{fig:enter-label}
\end{figure}
\subsection{MILP-Cov and MILP-Mov Performance}
After the static node placement, we then place the mobile nodes in the network grid. In Figure 4, the results of MILP-Mov algorithm is observed, which gives the movements taken to cover the full network area (cr = 1) for the network of $7\times7$ and increasing static and mobile nodes. The figure shows that total movements decrease by increasing the total static and mobile nodes.
Figures 5 and 6 show the behavior of MILP-Cov and MILP-Mov with different number of nodes. The MILP-Mov reduces the energy consumption by reducing the total movements for full coverage of network area (cr = 1). In Table 3, the effect of sensing and traveling range on area coverage using MILP-Cov is shown.

\begin{figure}
\centering
\includegraphics[width=0.7\linewidth]{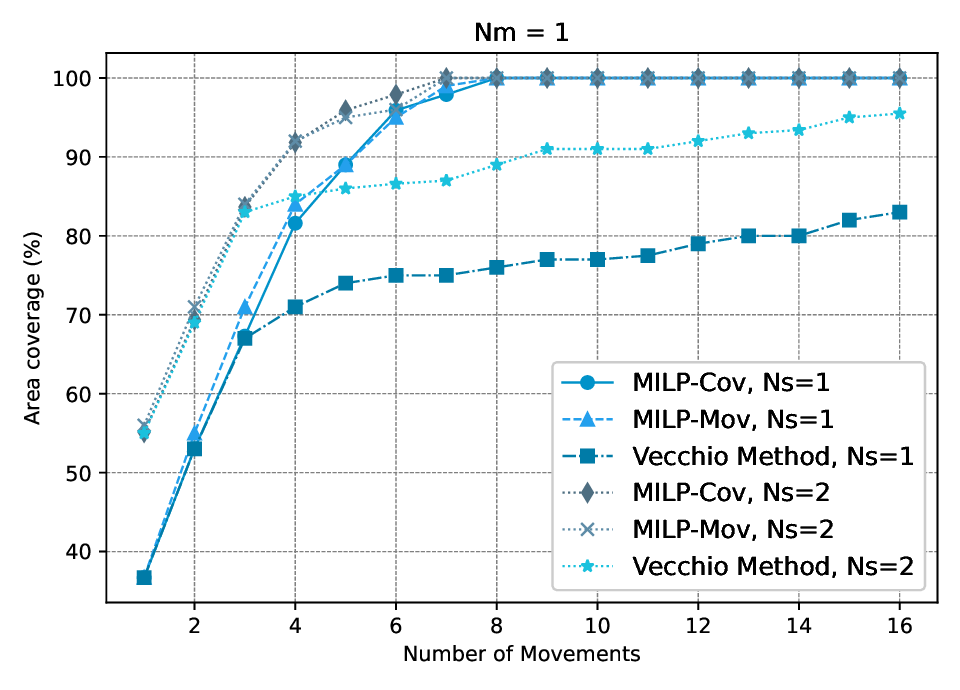}
\caption{Comparison of number of movements concerning area coverage with $N_m = 1$ with different number of nodes.}
\label{fig: 5(a)}
\end{figure}

\begin{figure}
\centering
\includegraphics[width=0.7\linewidth]{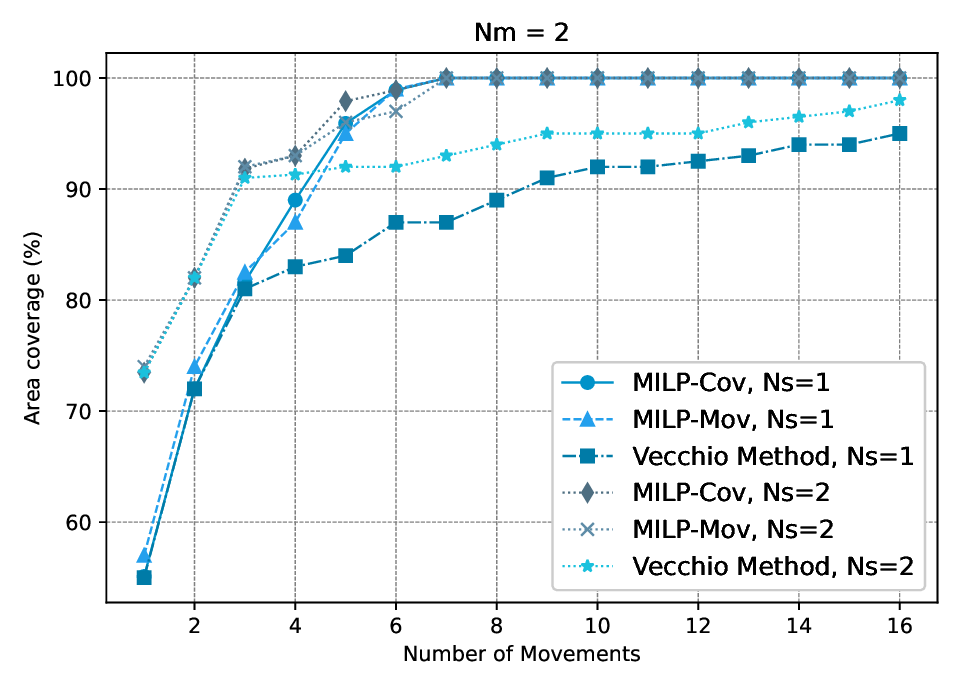}
\caption{Comparison of number of movements concerning area coverage with $N_m = 2$ with different number of nodes.}
\label{fig: 5(b)}
\end{figure}
 
\begin{table}[h!]
\centering
\caption{Sensing and traveling range on area coverage using MILP-Cov in a $7 \times 7$ network area with $N_s = 1$, $N_m = 2$, and $K_{max} = 2$.}
\begin{tabular}{|c|c|c|}
\hline
$r_s$ & $\rho_x = \rho_y$ & \textbf{Area Coverage (\%)} \\
\hline
& 1 & 75.51\\ & 2 & 81.63\\ 1& 3 & 83.67\\ & 4 & 83.67\\ & 5& 83.67\\
& 5 ($K_{max} = 3$)&95.91\\ 
\hline
\multicolumn{3}{c}{}\\
\end{tabular}
\label{tab:milp_cov}
\end{table}
\vspace{-1cm}
\subsection{Experimental Results for Energy Efficiency}
For an energy-efficient network, we experiment with it by setting up a Zigbee node as a transmitter, receiver, idle, and sleep mode to measure its power consumption. We connect the positive terminal of the multimeter to the power source of 5V and the negative to the $V_{in}$ of the Zigbee adapter. The resultant power is given below in Table 4.

\begin{table}[h!]
\centering
\caption{Results of Zigbee power consumption experiment with 5V power source.}
\begin{tabular}{|c|p{6cm}|p{5cm}|}
\hline
\multicolumn{1}{|c|}{\textbf{Mode for Experiment}} & 
\multicolumn{1}{c|}{\textbf{Resultant Current Consumption}} & 
\multicolumn{1}{c|}{\textbf{Power Consumption (at 5V)}} \\
\hline
\multicolumn{1}{|c|}{Transmitter Mode} & 
\multicolumn{1}{c|}{34.7 mA} & 
\multicolumn{1}{c|}{173.5 mW} \\
\hline
\multicolumn{1}{|c|}{Receiver Mode} & 
\multicolumn{1}{c|}{35.9 mA} & 
\multicolumn{1}{c|}{179.5 mW} \\
\hline
\multicolumn{1}{|c|}{Idle Mode} & 
\multicolumn{1}{c|}{35.1 mA} & 
\multicolumn{1}{c|}{175.5 mW} \\
\hline
\multicolumn{1}{|c|}{Sleep Mode} & 
\multicolumn{1}{c|}{0.1 mA} & 
\multicolumn{1}{c|}{0.5 mW} \\
\hline
\end{tabular}
\end{table}

\section{Conclusion}
The global expansion of internet of things (IoT) demands cost-effective solutions for power management, coverage, and resource constraints in wireless sensor networks (WSNs). Our mixed integer linear programming (MILP) framework addresses this through MILP-Static to optimize static placement up to 53.05\% compared to 33.42\% in random coverage. Next, we show that MILP-Cov enables 97.95\% coverage via mobile path planning, and MILP-Mov reducing movement costs by 40\%. Validated by simulations and power measurement, our approach minimizes unnecessary mobility while maximizing resource utilization in Zigbee-based networks. In future work, we plan to integrate AI-based energy-aware routing protocols with the MILP framework to further enhance network design, planning, and resource optimization. 

\bibliographystyle{ieeetr}
\bibliography{ref}
\end{document}